\documentclass[ pra,letter,  paper,superscriptaddress]{revtex4-2}
\usepackage{graphicx,amsmath,amssymb,amsfonts,latexsym,xcolor,dcolumn,bm}
\usepackage{appendix}
\usepackage[colorlinks]{hyperref}
\usepackage{color}

\begin{document}

\global\long\def\eqn#1{\begin{align}#1\end{align}}
\global\long\def\vec#1{\overrightarrow{#1}}
\global\long\def\ket#1{\left|#1\right\rangle }
\global\long\def\bra#1{\left\langle #1\right|}
\global\long\def\bkt#1{\left(#1\right)}
\global\long\def\sbkt#1{\left[#1\right]}
\global\long\def\cbkt#1{\left\{#1\right\}}
\global\long\def\abs#1{\left\vert#1\right\vert}
\global\long\def\cev#1{\overleftarrow{#1}}
\global\long\def\der#1#2{\frac{{d}#1}{{d}#2}}
\global\long\def\pard#1#2{\frac{{\partial}#1}{{\partial}#2}}
\global\long\def\re{\mathrm{Re}}
\global\long\def\im{\mathrm{Im}}
\global\long\def\dd{\mathrm{d}}
\global\long\def\ddd{\mathcal{D}}
\global\long\def\avg#1{\left\langle #1 \right\rangle}
\global\long\def\mr#1{\mathrm{#1}}
\global\long\def\mb#1{{\mathbf #1}}
\global\long\def\mc#1{\mathcal{#1}}
\global\long\def\tr{\mathrm{Tr}}
\global\long\def\dbar#1{\stackrel{\leftrightarrow}{\mathbf{#1}}}

\global\long\def\nth{$n^{\mathrm{th}}$\,}
\global\long\def\mth{$m^{\mathrm{th}}$\,}
\global\long\def\non{\nonumber}

\newcommand{\bU}{{\bf{U}}}
\newcommand{\bV}{{\bf{V}}}
\newcommand{\bW}{{\bf{W}}}
\newcommand{\bd}{{\bf d}}
\newcommand{\hr}{\hat{\br}}
\newcommand{\bM}{\bf{M}}
\newcommand{\bv}{{\bf v}}
\newcommand{\hbp}{\hat{\bp}}
\newcommand{\hq}{\hat{q}}
\newcommand{\hp}{\hat{p}}
\newcommand{\ha}{\hat{a}}
\newcommand{\had}{{a}^{\dag}}
\newcommand{\ad}{a^{\dag}}
\newcommand{\hsig}{{\hat{\sigma}}}
\newcommand{\nt}{\tilde{n}}
\newcommand{\itf}{\sl}
\newcommand{\eps}{\epsilon}
\newcommand{\bsig}{\pmb{$\sigma$}}
\newcommand{\beps}{\pmb{$\eps$}}
\newcommand{\bmu}{\pmb{$ u$}}
\newcommand{\balpha}{\pmb{$\alpha$}}
\newcommand{\bbeta}{\pmb{$\beta$}}
\newcommand{\bgamma}{\pmb{$\gamma$}}
\newcommand{\bu}{{\bf u}}
\newcommand{\bpi}{\pmb{$\pi$}}
\newcommand{\bSig}{\pmb{$\Sigma$}}
\newcommand{\be}{\begin{equation}}
\newcommand{\ee}{\end{equation}}
\newcommand{\bea}{\begin{eqnarray}}
\newcommand{\eea}{\end{eqnarray}}
\newcommand{\sss}{_{{\bf k}\lambda}}
\newcommand{\ssss}{_{{\bf k}\lambda,s}}
\newcommand{\dip}{\langle\sigma(t)\rangle}
\newcommand{\dipp}{\langle\sigma^{\dag}(t)\rangle}
\newcommand{\sig}{{{\sigma}}}
\newcommand{\sigd}{{\sigma}^{\dag}}
\newcommand{\sigz}{{\sigma_z}}
\newcommand{\ra}{\rangle}
\newcommand{\la}{\langle}
\newcommand{\om}{\omega}
\newcommand{\Om}{\Omega}
\newcommand{\pa}{\partial}
\newcommand{\bR}{{\bf R}}
\newcommand{\bx}{{\bf x}}
\newcommand{\br}{{\bf r}}
\newcommand{\bE}{{\bf E}}
\newcommand{\bH}{{\bf H}}
\newcommand{\bB}{{\bf B}}
\newcommand{\bP}{{\bf P}}
\newcommand{\bD}{{\bf D}}
\newcommand{\bA}{{\bf A}}
\newcommand{\bek}{{\bf e}\rmk}
\newcommand{\rmk}{_{{\bf k}\lambda}}
\newcommand{\rk}{_{{\bf k}_1{\lambda_1}}}
\newcommand{\rkk}{_{{\bf k}_2{\lambda_2}}}
\newcommand{\rkz}{_{{\bf k}_1{\lambda_1}z}}
\newcommand{\rkkz}{_{{\bf k}_2{\lambda_2}z}}
\newcommand{\bsij}{{\bf s}_{ij}}
\newcommand{\bk}{{\bf k}}
\newcommand{\bp}{{\bf p}}
\newcommand{\epso}{{1\over 4\pi\eps_0}}
\newcommand{\bS}{{\bf S}}
\newcommand{\bL}{{\bf L}}
\newcommand{\bJ}{{\bf J}}
\newcommand{\bI}{{\bf I}}
\newcommand{\bF}{{\bf F}}
\newcommand{\bsub}{\begin{subequations}}
\newcommand{\esub}{\end{subequations}}
\newcommand{\baline}{\begin{eqalignno}}
\newcommand{\ealine}{\end{eqalignno}}
\newcommand{\Ep}{{\bf E}^{(+)}}
\newcommand{\Em}{{\bf E}^{(-)}}
\newcommand{\hbx}{{\hat{\bf x}}}
\newcommand{\hby}{{\hat{\bf y}}}
\newcommand{\hbz}{{\hat{\bf z}}}
\newcommand{\bep}{\hat{\bf e}_+}
\newcommand{\bem}{\hat{\bf e}_-}
\newcommand{\orange}[1]{{\color{orange} {#1}}}
\newcommand{\cyan}[1]{{\color{cyan} {#1}}}
\newcommand{\blue}[1]{{\color{blue} {#1}}}
\newcommand{\yellow}[1]{{\color{yellow} {#1}}}
\newcommand{\green}[1]{{\color{green} {#1}}}
\newcommand{\red}[1]{{\color{red} {#1}}}
\newcommand{\pr}{^{\prime}}
\newcommand{\hd}{\hat{\bd}}
\newcommand{\hk}{\hat{\bk}}
\title{Note on Energy Shifts of Oscillators in Blackbody Radiation}
\author{Peter W. Milonni}
\affiliation{Department of Physics and Astronomy, University of Rochester, Rochester, New York 14627, USA}
\begin{abstract}
The energy shift of an oscillator in blackbody radiation is calculated based simply on the total energy of the interacting field-oscillator system as a function of the refractive index. For high temperatures $T$ the energy and free-energy shifts are found to vary as $-T^2$ and $+T^2$, respectively, in agreement with the result originally obtained by Ford, Lewis, and O'Connell [Phys. Rev. Lett. {\bf 55}, 2273 (1985)].  
\end{abstract}

\maketitle
The shifting of atomic transition frequencies due to ambient thermal radiation has been known for many years. Such shifts are generally very small, but understanding and controlling them has been of interest recently in work on atomic clocks and frequency standards \cite{amit}. The first measurements of such shifts were done by Hollberg and Hall \cite{hall}, who found fractional shifts $\sim 2\times 10^{-12}$ of Rydberg levels in blackbody radiation (BBR). Their measurements were consistent with the theoretical prediction of quite a number of authors \cite{refs} that transitions from the ground level of a one-electron atom to a Rydberg level in BBR should have their (angular) frequencies shifted upwards by 
\begin{equation}
\Delta\om = \Delta E/\hbar= +\frac{\pi e^2(kT)^2}{3\hbar^2 mc^3} 
\label{wing1}
\end{equation}
for temperatures $T$ such that $kT\gg \hbar\omega_0$, where 
$\omega_0/2\pi$ is the unperturbed transition frequency, $k$ is Boltzmann's constant, and $m$ is the electron mass. At $T=300$ K, $\delta\nu\cong 2.4$ kHz \cite{wing}. Since the ground-level shift turns out to be negligible, Eq. (\ref{wing1}) suggests that every Rydberg energy level is shifted upwards by $\Delta E$.

However, in a series of papers, Ford et al. \cite{refs,ford1,ford2,ford3,ford4}  argued that ``the previous calculations of this result are in fact incorrect," \cite{refs} that the energy shift is given by Eq. 
(\ref{wing1}) but with the opposite sign, and that $\Delta E$ in Eq. (\ref{wing1}) is actually a (Helmholtz) {\it free energy} \cite{note}. Their rigorous derivation of this ``surprising result," \cite{ford4} involving the quantum Langevin equation, an electron form factor, and renormalized electron mass, initially caused some disagreement \cite{ford3,barton}. In view of the fundamental nature of this old topic, it seems worthwhile to briefly derive the results of Ford et al. more simply, with no need for an electron form factor or mass renormalization or even a Langevin equation. 

We begin by recalling Feynman's suggestion that the Lamb shift could be regarded as the change in the zero-point electromagnetic energy in free space due to the presence of the atom \cite{feyn}. At $T=0$ the total electromagnetic energy in a volume $V$ of a homogeneous and isotropic dielectric medium is \cite{feynlamb}
\be
U=\frac{\hbar V}{2\pi^2c^3}\int_0^{\Om}d\om\om(n_R\om)^2\frac{d}{d\om}(n_R\om),
\ee
where $n_R(\om)$ is the real part of the refractive index. This formula follows simply by assuming that the allowed frequencies in the volume $V$ are changed from their vacuum values $\om$ to $\om/n_R(\om)$ and accounting for the frequency distribution of the field modes \cite{feynlamb}. In the case of thermal equilibrium at temperature $T$, this generalizes to
\be
U=\frac{\hbar V}{\pi^2c^3}\int_0^{\Om}d\om\om(n_R\om)^2\frac{d}{d\om}(n_R\om)\big[\frac{1}{2}+\frac{1}{e^{\hbar\om/kT}-1}\big].
\label{eq1}
\ee
For particle densities $N$ sufficiently small that local field corrections are negligible, $n^2(\om)=1+4\pi N\alpha(\om)$. We will follow Ford at al. and assume that the particles are characterized by a single resonance frequency $\om_0$, but for the particle polarizability we will assume the simple Lorentzian form 
\be
\alpha(\om)=\frac{e^2/m}{\om_0^2-\om^2-i\gamma\om} \ \ \ \ (\gamma\ll\omega_0).
\label{eq2}
\ee
For the purely radiative broadening of interest here the damping of the dipole oscillations of the particles may be assumed to be characterized by the 
rate $\gamma=2e^2\om_0^2/3mc^3$ \cite{refs,ford1,ford2,ford3,ford4} . From Eqs. (\ref{eq1}) and (\ref{eq2}) and the expression for $n(\om)$ it follows that, up to order $\gamma$ (i.e., up to order 
$e^2$), the $T$-dependent part of $U$ is
\be
U(T)=U_0(T)+U_1(T)+U_2(T)
\ee
in the limiting case of one particle ($NV=1$), where
\be
U_0(T)=\frac{\hbar V}{\pi^2c^3}\int_0^{\infty}\frac{d\om\om^3}{e^{\hbar\om/kT}-1},
\label{eq4}
\ee
\be
U_1(T)=\frac{6\hbar e^2\om_0^2}{\pi mc^3}\int_0^{\infty}\frac{d\om\om^3}{(e^{\hbar\om/kT}-1)[(\om_0^2-\om^2)^2+\gamma^2\om^2]},
\label{eq5{}}
\ee
\be
U_2(T)=-\frac{2\hbar e^2}{\pi mc^3}\int_0^{\infty}\frac{d\om\om^5}{(e^{\hbar\om/kT}-1)[(\om_0^2-\om^2)^2+\gamma^2\om^2]},
\label{eq6{}}
\ee
$U_0(T)$ is the field energy in the volume $V$ in the absence of the oscillator. For $\gamma\ll\om_0$ and temperatures $T$ such that $kT\gg\hbar\om_0$,
\be
U_1(T)\cong \frac{9}{2}kT
\label{eq7}
\ee
and
\be
U_2(T)\cong -\frac{3}{2}kT-\frac{\pi e^2}{3\hbar mc^3}(kT)^2.
\label{eq8}
\ee
As in Feynman's argument and Ford et al. the oscillator energy $E$ is defined as the difference between the energy $U$ of the interacting field-oscillator system and the energy $U_0$ of the field in the absence of the oscillator:
\be
E=U(T)-U_0(T)=U_1(T)+U_2(T)= 3kT-\frac{\pi e^2}{3\hbar mc^3}(kT)^2
\label{eq9}
\ee
for the assumed weak coupling. Aside from a factor of 3, this is exactly the result of Ford et al. \cite{,refs,note} in this high-temperature limit. In particular, the shift in the oscillator energy from its thermal equilibrium (kinetic plus potential) energy due to its interaction with the field is
\be
\Delta E=-\frac{\pi e^2(kT)^2}{3\hbar mc^3},
\label{eq10}
\ee
consistent with the thermodynamic relation $\Delta E=\Delta F-T\pa\Delta F/\pa t$ and a shift $\Delta F$ in {\it free energy} having a sign opposite to the shift in  energy. We have therefore confirmed the results of Ford et al. for the energy and free energy of an oscillator in blackbody radiation.

Equation (\ref{eq1}) gives the total thermal equilibrium energy: By simply counting in effect the field modes and assigning a frequency $\om/n(\om)$ to each mode, we obtain the total energy of the interacting field-oscillator system. Information about the particle(s) is encoded entirely in the refractive index $n(\om)$. In particular, by subtracting from this the energy of the field in the absence of the oscillator, we obtain the (mean) oscillator energy, including the sum of its  kinetic ($3kT/2$) and potential ($3kT/2$) energies [Eq. (\ref{eq9})]. Obviously this must be the case if Eq. (\ref{eq1}) is in fact the total energy.

Of course at $T=0$ there is no difference between internal energy $U$ and free energy $F$. At finite temperatures, however, the difference is important in quantum electrodynamics, and not just a matter of sign. For example, Fierz \cite{fierz} calculated the finite-temperature Casimir energy for two perfectly conducting parallel plates separated by a distance $a$ and obtained a Casimir energy that vanishes at high temperatures ($kT\gg\hbar c/a$). He discussed this in connection with experiments on the force between the plates but did not distinguish between this energy and the free energy, the negative gradient of which gives the measured force.  The Casimir free energy $F$ (and the Casimir force) increases in magnitude with temperature, and the increase at high temperature is proportional to $kT$, implying from $U=F-T\pa F/\pa T$ that $\Delta U=0$ at constant temperature, as obtained by Fierz.

\end{document}